# Designing for Employee Voice

Dinislam Abdulgalimov*, Reuben Kirkham*, James Nicholson^, Vasilis Vlachokyriakos~,
Pam Briggs^, Patrick Olivier*
*Monash University, Australia; ^Northumbria University, UK; ~Newcastle University, UK.
dinislam.abdulgalimov@monash.edu

**ABSTRACT**
Employee voice and workplace democracy have a positive impact on employee wellbeing and the performance of organizations. In this paper, we conducted interviews with employees to identify facilitators and inhibitors for voice within the workplace and a corresponding set of appropriate qualities: *Civility*, *Validity*, *Safety* and *Egalitarianism*. We then operationalised these qualities as a set of design goals – *Assured Anonymity*, *Constructive Moderation*, *Adequate Slowness* and *Controlled Access* – in the design and development of a secure anonymous employee voice system. Our novel take on the Enterprise Social Network aims to foster good citizenship whilst also promoting frank yet constructive discussion. We reflect on a two-week deployment of our system, the diverse range of candid discussions that emerged around important workplace issues and the potential for change within the host organization. We conclude by reflecting on the ways in which our approach shaped discourse and supported the creation of a trusted environment for employee voice.

**CSS Concepts**
•**Human-centered computing~Human computer interaction (HCI)**; *Collaborative and social computing*;

**INTRODUCTION**
Work cultures and environments have an enormous impact on employee wellbeing. Working-aged adults spend an average of 40-47 hours a week within the physical workplace [2, 62]. Accordingly, personal dissatisfaction with the workplace environment can adversely affect individual quality of life [11,16,30]. Much of the dissatisfaction with workplaces stems from hierarchical management structures and disparities in power within the organization [22]. One approach known to have a positive impact on (hierarchical) workplaces is the creation of a culture and associated organizational systems that promote *employee voice* [29]. Organizations that implement mechanisms to capture and respond to employee voice, have been shown to benefit from increased staff retention and satisfaction, improved reputation, and a better understanding of their own operations [20,52,77]. These 'listening' organizations are able to take advantage of the accumulated wisdom within the workplace in order to show enhanced decision making [77], whilst employees benefit from improved mental health, better work-life balance, workplace satisfaction, and a sense of 'being heard' [11,16,30].

Given the complexity of industrial relations, and the amount of different workplace-related issues that emerge across the globe – from workplace bullying [69] and labor law violations [18, 46] to discrimination and harassment [65] – it is unsurprising that there are serious challenges to the promotion of employee voice. These challenges can be expressed as two important trust issues. On the one hand, employees need to be persuaded that they can express their views honestly, constructively and without risk of reprisal. On the other hand, employers need to be convinced that the concerns raised are properly representative of the situation on the ground in their organization. The result is that employees generally prefer anonymity, informality, and collectivism but employers prefer formal consultation mechanisms that reflect existing organizational structures [22]. More recently, Enterprise Social Networks (ESNs) have been introduced in many workplaces with the stated intention of furnishing employees with the opportunity to engage directly with fellow workers [15], but recent evidence suggests that these are seen by employees as either channels for self-promotion or vehicles for employee surveillance [44].

In this paper we establish a set of core design goals for a novel Enterprise Social Network that can better facilitate employee voice and thus promote the establishment of horizontal peer-to-peer communication about workplace issues within an organization. Based on interviews and workshops, we identified that certain characteristics of discourse – *Civility*, *Validity*, *Safety* and *Egalitarianism* – serve as *appropriate qualities* for workplace conversations that genuinely constitute employee voice. We then operationalised these qualities as a set of design goals – *Assured Anonymity*, *Constructive Moderation*, *Adequate Slowness* and *Controlled Access* – in the design and development of a secure anonymous employee voice system. We deployed our bespoke anonymous ESN to more than 600 staff members within a sub-unit of a UK University, and assessed the extent to which our system met the design goals and supported high quality, civil, valid, safe and egalitarian discourse. Our contribution is three-fold: *(i)* we identify facilitators and inhibitors for voice within the workplace and a corresponding set of appropriate qualities; *(ii)* we operationalize these qualities as both a set of design goals for an employee voice system, and an ESN-type system that we designed and fully implemented; and *(iii)* from a real-world deployment and evaluation of our system we show how a combination of trusted anonymity and 'slowed' moderation can foster employee voice.

## BACKGROUND

### Employee Voice

Employee voice was originally defined as 'providing workers as a group with a means of communicating with management' [29], but was later expanded to include participation in decision making, engagement in workplace discussions and the ability to express opinions freely without fear of repercussions [77]. It is typically seen as an important empowerment mechanism and a facilitator of bottom-up participatory planning within an organization. Over a number of years the concept of employee voice has become more refined, leading to the identification of four goals: *(i)* the articulation of personal dissatisfaction with processes; *(ii)* the expression of collective decisions and thoughts; *(iii)* the chance to contribute towards relevant management decisions; and *(iv)* the display of mutuality in employer-employee relationships [50]. The notion of *workplace democracy*, an element of Participatory Design (PD), has similar goals to that of employee voice. Workplace democracy involves providing employees with the opportunity to influence their work environment through their participation in decision making processes [7]. For example, in Norway, such projects have led to legislative changes that empower workers as well as policies that promote worker engagement as a means to increase productivity and efficiency [28,67]. Similarly, the Scandinavian Participatory Design school argues for a balancing out workplace power dynamics and putting greater focus on workers' interests and 'support[ing] the development of their resources towards democracy at work' [27].

There are both formal and informal systems and approaches for employee voice. Formal voice is typically captured using pre-existing management and communication structures and timelines [47]. By contrast, informal voice allows for the expression of employee opinion at any time, and not necessarily as a response to a management-led query [41]. Employees use both formal and informal mechanisms [22] but usually prefer informal systems [55]. In a study of Australian workplaces, it was found that formal channels were more effective for groups of workers, whilst informal routes are preferred for issues involving individual rights and preferences [72].

If properly captured and actioned by an organization, promoting effective employee voice brings benefits for employers and employees alike. For employers, employee voice is in collective decision-making is often found to be most effective [52]. Further, it facilitates the early detection and simpler resolution of problems within the organization [26,51]. For employees, employee voice is known to enhance work satisfaction [20,26], has a positive impact upon mental health [3,30,31], and contributes towards employees feeling valued within the workplace [52], which in turn can improve employee retention [20].

Yet in practice, it is difficult to create and sustain forms of meaningful communication and collaboration between workers. Workers often believe that frank expression of dissatisfaction, or potentially controversial suggestions for new ways of working, will result in negative repercussions or retaliation, even when offered in good faith. The most prominent concerns are that job security or career progression will be adversely affected [9,21]. This applies even when the 'contribution' is intended to benefit the organization [19]. Likewise, individuals are less likely to contribute to a group discussion where there is a power imbalance [36] and 'powerful' individuals are often found to make implicit threats about 'rocking the boat' [9,23]. Accordingly, there are significant barriers within organizational ecosystems that can limit employee contributions beyond those strictly aligned with workers' roles [59]. There is a second less obvious concern. Speaking out constitutes an investment of time and energy, and on occasion emotional investment, especially if a concern is something that an employee is particularly passionate about [33]. Yet, when workers speak out, it is easy for managers to discount their contributions. This can lead to a vicious circle. If workers believe that their contributions will stay unaddressed then they will view the process as pointless and simply not initiate or support actions [24]. The provision of a meaningful and respected structure for cooperative employee action is therefore important to ensure that there is no 'hollow shell' or pretence at responding to actions [71].

### Enterprise Social Networks (ESN)

Enterprise Social Networks (ESNs) are intended to promote peer-to-peer engagement between employees and access knowledge 'bottom-up', with pertinent examples being IBM's (now defunct) SocialBlue, and the more widespread Slack and Yammer. ESNs are modelled on popular social networking sites such as Facebook and Twitter (indeed, Facebook has its own workplace solution) and offer employees the opportunity to form work groups comprising individuals with shared interests. ESNs such as these are seen as valuable by employees for the purposes of following others, obtaining news [56] and seeking information from expert peers [15] as well as continuing professional conversations outside of regular work hours [75]. Previous work has examined ESNs as a tool to encourage innovation and creativity [61] as well as to improve team collaborations [49], although employees must be able to see the potential for knowledge exchange in order to accept such tools [5]. Additionally, there is evidence suggesting that ESNs can be used for organizational problem solving, albeit in the form of questions and answers [15].

The flipside of this is that ESNs can also function as tools for management to monitor employees' perceptions of the organization [64]. Such uses can backfire with workers perceiving a loss of power and privacy, which inhibits ESN adoption [44]. An evaluation of social media guidelines for ESNs found that personal transparency – such as the ability to identify the identity of individuals posting – was one of the top factors negatively affecting adoption [42] and a key reason why employees often fail to engage with such tools [56]. This points to a need for *anonymous communication* in the workplace.

## DESIGN

The goal of our study was to respond to the many limitations of existing approaches to supporting employee voice by focusing upon the social value of anonymity. However, from the outset we were very mindful of likely reservations about anonymous systems and of the value of understanding workers' experiences of speaking up in the past. We therefore conducted a series of semi-structured interviews of University employees (the category of employees that comprise our participants) with a view to identifying potential barriers and facilitators, as well as their attitudes towards digital facilitation for 'speaking up' and their concerns about anonymous discourse. Our analysis of these interviews helped us construct a set of design goals for the design and development of a digital system to capture employee voice, described in the following section.

### Domain Exploration and Goals Formulation

Fourteen semi-structured interviews were conducted across two universities (the workplaces of members of the research team), covering a total of nine different academic departments and comprising 7 support and 7 academic staff. The participants were all recruited via email (through internal local trade union mailing lists). The interviews were divided into two sections that addressed: *(i)* experiences of speaking up, or witnessing someone else doing so, and generating change within a current or previous organization; and *(ii)* ideas about how technology might be used to facilitate employee voice in the workplace.

### Communication Qualities

We conducted a qualitative thematic analysis of part *(i)* of the interview data (experiences of speaking up, etc.) by adhering to what Braun & Clarke [8] describe as a "more detailed and nuanced account of one particular theme, or group of themes, within the data". Specifically, we employed a 'theoretical' thematic analysis that was informed by previous studies of employee voice, in particular, personal and organizational *facilitators* and *inhibitors*, that impact on decisions about whether or not to speak-up.

*Personal inhibitors.* These are the factors that inhibit our participants speaking out or engaging in discussions about workplace issues. They included: *(i)* uncertainty about how to raise an issue; *(ii)* a lack of trust in the organization's ability or desire to address the issue, and the belief that any issues raised would be ignored; *(iii)* the potential to be perceived as a problematic and uncooperative; *(v)* being seen as the cause of disruption and uncomfortable atmosphere; *(v)* fear of sanctions or reprisals.

*Personal facilitators:* Our participants recognized that their organization could sanction those individuals prepared to speak out, either directly or indirectly (e.g. via subsequent job progression). The need for a safe and trusted environment for speaking-up was made clear. Another important aspect was the need for parity between all engaging entities (e.g. employees, managers, owners). The desire was for a system that should be open to all and that should promote listening.

*Organizational facilitators.* These are the factors that increase the likelihood that employees would pursue coordinated action. They included: *(i)* approachable and transformational managers; *(ii)* the existence of clear bottom-up communication channels through which to express views; and *(iii)* an organization's track record of responding meaningfully to issues raised by employees.

*Organizational inhibitors.* These are the factors reported by our participants, that decrease the likelihood that employees would pursue coordinated action. They included: *(i)* transactional and overly formal managers; *(ii)* poor or non-existent communication channels for employees to express of their views; and *(iii)* an organization's track record of ignoring staff dissatisfaction (e.g. by repeating mistakes).

By considering the phenomena of communication more generally, as it pertains to aspirations for employee voice, we can derive a set of communication *qualities* for workplace conversation as an analytical abstraction of our themes relating to facilitators and inhibitors:

*Civility*: a conversation should be polite and respectful. Here we adhere to Brown and Levinson's definition [10] and the need to maintain a positive, constructive approach to the community you interact with.

*Safety*: a conversation should be conducted in a safe and un-threatening manner, in order to promote a more open and frank discussion.

*Egalitarianism*: the conversation content, organization and flow should be indifferent to the employees' position or level in the organization.

*Validity*: conversations should be genuine, truthful, relevant and authentic.

### Design Goals

Having conceived these communication *qualities* for workplace conversation, our next step was to analyse our part *(ii)* of the interviews, namely participants' ideas about how technology might be used to facilitate employee voice in their workplace. This iteration our thematic analysis of the interview data took the perspective of participants' understandings of the pragmatics of realising the communication qualities (derived in the first iteration of the analysis) to develop themes that we operationalize as a set of design goals. The resulting goals – *Assured Anonymity*, *Constructive Moderation*, *Adequate Slowness* and *Controlled Access* – defined the primary scope of our ESN-type employee voice system.

***Assured Anonymity:*** *Assured Anonymity* operationalizes aspects of *Egalitarianism* and *Safety* [39]. Anonymity effectively eliminates any hierarchical structure within the group and facilitates equal contribution whilst protecting members from any threat of reprisal. It is unsurprising that all of the participants that we canvassed were insistent that *anonymity* was a key component in an employee voice system, and equally, that employees must have *confidence* in the efficacy of the system in protecting that anonymity:

> I12: "I think anonymity is really an important part of this… if there is not confidence that there is a robust, anonymous system I don't think it would be worth doing."

However, a minority of participants argued that anonymity should be limited, for instance, to a 'trusted authority or *"superuser"* (I6), because a purely anonymous system would be subject to *"flaming"* (I2). This raises a dilemma as to the relative benefits and costs of supporting anonymity for such services. Notable systems such as *Secret* (now defunct*),* that allowed people to share messages anonymously, was closed down by its founder who cited "ethical issues" [1]. An earlier service, *Juicy Campus*, that allowed users for vote for "juiciest" anonymously posted gossip about college campuses, was closed after failed attempts to encourage users to be less 'cruel' [34, 45]. A similar fate befell *Yik-Yak* [68]*,* which in its time was the predominant (and most actively studied) anonymous system; as well as *Blind*, a digital service to facilitate anonymous sharing amongst employees [76]. Notably, *Blind* required users to register with their work email address or by providing access to their LinkedIn profile, and depended on *reactive moderation*, meaning that a post was only removed if enough users flagged it.

***Constructive Moderation:*** *Constructive Moderation* operationalizes aspects of *Civility*, *Egalitarianism* and *Safety*. Our interviewees were adamant that anonymity must be assured, but also argued that people must be protected against abuse. Taken together, these suggested the need for a careful moderation system. Moderation is a process by which posts are edited or removed in order to maintain specified standards of behaviour and expression in interactions. It is well known that anonymous online communities can have the effect of eroding traditional self-censorship, with users being willing to over-disclose [38] and/or engage in frank conversations that exceed professional norms [25]. Such online disinhibition can lead to toxic behaviour, such as directly insulting or flaming other participants [66]. Moderation, therefore, acts to protect users from themselves (by including checks against inadvertent personal disclosures that would pierce the cloak of anonymity) and also to protect against others (from abuse and/or inappropriate disclosure). Moderation also helps to ensure that online discussions are genuinely productive, especially in the setting of online debates [35, 73].

Effective moderation is not easy. The most obvious challenge is the resource burden, which can be higher in anonymous forums as they are associated with higher levels of posting activity and more inflammatory posts [40, 60]. To reduce this burden, many systems rely upon reactive moderation, for example, using reputation scores to control the ways that certain users are allowed to post [43] or using automation [14] to manage the moderation burden. Reactive moderation is efficient, but its limitations are clear: we can remove offensive posts, but in many circumstances we are closing the 'barn door' too late [70]. For proactive moderation (i.e. control of content prior to release) submitted posts must each be manually reviewed. This has been shown to reducing bullying and negativity [6], but is resource intensive and usually relies on large numbers of volunteers prepared to make significant time commitments. There is also the question of how moderation processes are governed. There is a fine line between moderation and censorship. Indeed, the fear of censorship can have a chilling effect, reducing the quality of conversations and debate [54]. Moreover, poor moderation can limit the participation of minorities, especially where no steps are taken to address the risk of stereotypically defined and driven topics [58]. Moderation therefore requires skill, integrity, and the ability to command user respect. Given the levels of required skills and training [17], most human moderation will be somewhat imperfect, and any system will need to take this into account, for example through group review (i.e. distributing responsibility across multiple moderators).

***Adequate Slowness***: *Adequate Slowness* operationalizes *Civility* and *Egalitarianism*. We have already noted that one potential by-product of anonymity can be incivility, which can be filtered using careful and robust moderation. However, flaming often takes the form of a swift and angry response to a comment. By slowing system post publication, thereby removing the opportunity to make an immediate response, we can at the same time reduce the likelihood of incivility and make the moderation burden more manageable. In our eventual design we slowed the system by limiting the publication of posts and comments to certain times of day. This also induced a form of synchrony to what is an intrinsically asynchronous system, whereby people would tend to view the system and make contributions at those particular times of day. This brings a particular cadence to workplace communication which may be worth exploring further. The *slowness* of interactions also means that users are encouraged to think on their responses (knowing they will only be published hours or more later) and mitigate against short outbursts in favour of more reasoned and reflective posts and comments.

***Controlled Access:*** From both our interviews and existing research on employee voice it is clear that the involvement of outsiders who have no legitimate interest in the workplace contexts is highly undesirable. By providing a controlled access mechanism, we can ensure the *Validity* of discussions as these can only be made by genuine employees. With the important provision that no division should be made between employees in order to protect discourse *Egalitarianism.* The majority (10/14) of participants in our initial interviews took the view that *all staff members* should be included:

> I1: "… without them our jobs would become impossible, or very unpleasant. So, technical support staff, cleaners, porters, everybody, has a stake in this university. And everybody that has a stake should have a say."

Our participants emphasized that discussions should ultimately lead to actions that could bring about meaningful change but were unsure about appropriate feedback mechanisms. It was considered important to have managers participate in an ICT-driven discussion for it to be widely accepted. In other words, this discussion should have extensive 'soft power', such that it could not be ignored [12].

## CASE STUDY

A prototype system, NINEtoFIVE, was designed and implemented to realize our design goals for the facilitation of employee voice. This was then deployed in a two-week trial, based in a university academic department of 600 staff and over 100 PhD students. Before the deployment, we conducted further workshops where we tested (and refined) the system and its usability in practice.

***The Deployment Context***: We chose to launch a two-week trial of NINEtoFIVE immediately following the move of an entire academic department into a new building on a separate campus. We teamed up with the academic trade union for the deployment of NINEtoFIVE, who adopted our system as an official initiative. While NINEtoFIVE was open to all staff and research students within the department, the initial advertising of the platform was directed towards union members with the option to invite non-union colleagues to engage with the system. One emphasis of the platform was therefore to discuss positive and negative aspects of the new building – although users were allowed to make posts on any topic as long as they were relevant to the University.

***Promoting NINEtoFIVE***: The initial advertisement of the system consisted of two emails sent by the local president of the union (to all union members within the department) and the Postgraduate Research Students' representative (to eligible students) early in the morning. The system was configured such that any member of the department with a valid university email address had token-based access (see the next subsection: 'Accessing NINEtoFIVE'). However, given that NINEtoFIVE was an initiative of the trade union (the trusted broker), it was initially promoted through a private email list of union members in the department, who were encouraged to forward promotional material to non-union colleagues. As the deployment progressed, we also distributed physical posters around the department (including summaries of discussions from the system) and sent supplementary promotional emails using departmental mailing lists. The emails contained a link to the NINEtoFIVE and described the functionality of the system. Another key element of our design was that neither union members nor non-union staff should feel pressured to participate. Therefore, not only was it impossible for us to know who took part, all communication emails emphasized the anonymous and the voluntary nature of participation.

***Accessing NINEtoFIVE***: Considerable care was taken to ensure that NINEtoFIVE operated as a convincingly secure platform where our participants could trust in the privacy of the system (the *Safety* principle). To this end, we used a 'token' based system, where users entered their work email for verification on the website (the *Validity* principle) and received a 'one-time-link' to their registered email address that would enable them to access the system. The link was not associated in any way with the email address – email was simply the method we chose for delivering the link to users. To this end, the system had no way of knowing who clicked on the link. For those who were concerned that their employer would know that they were regularly using the system, they were able to switch their work email address to a private email address to receive the tokens.

***Publishing Posts and Comments on NINEtoFIVE***: Anyone who used a 'link' to access the system could submit a new post (under one of the built-in categories within the system, eventually creating a new thread) or comment (on an existing post) to the moderators (see **Figure 1**). Text-based communication was used to address privacy concerns, given that videos and images are difficult to anonymize, as well as allowing moderators to respond proportionately to breaches of anonymity by users (e.g. through minor edits to posts). The moderators had the responsibility for either: *(i)* publishing the post or comment 'as is'; *(ii)* editing the post or comment; or *(iii)* rejecting the post or comment, in which case it would be permanently deleted from the system. The system would publish approved posts and comments in batches, at both 9am and 5pm, following formal moderation meetings prior to these times. The purpose of doing so was: *(i)* to improve anonymization, eliminating links between potentially observable interactions with the system and posts appearing(*safety*); *(ii)* to reduce distraction during working hours second by publishing on first and last working hours only; *(iii)* to decrease the pace of interaction, as a means of inhibiting flaming (c*ivility*) and preventing hijacking of discussions by more active users (e*galitarianism);* and *(iv)* in response to the pragmatics of moderation, in that face-to-face moderation meetings (used in our deployment) would only have to happen twice per day. Users could, however, vote posts up or down as a means of influencing discussion. As the system was anonymous, random usernames were generated for comments, which were tied to a single use of a 'login link'. These names were intended to have a ludic quality (e.g. '*complex chestnut sheep'*), as well as enabling easy references to what a previous poster had said within a discussion thread to support discourse flow. We also indicated the moderation session in which the post was approved (i.e. '*dd/mm/yy AM'*) to enable our participants to more easily navigate a given thread.

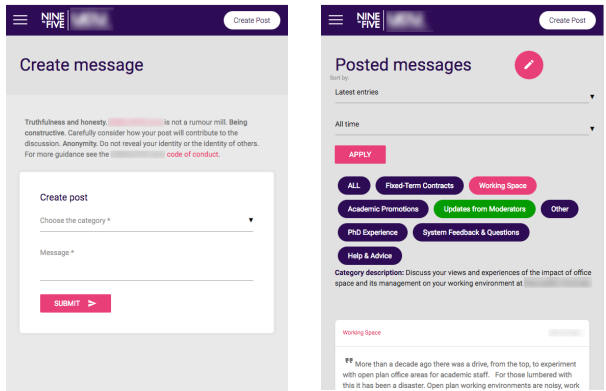

**Figure 1. Post creation & browsing in NINEtoFIVE.**

*Browsing NINEtoFIVE*: NINEtoFIVE contained a number of features to enable passive users (those who do not submit posts or comments) to interact with the system, as well as to identify posts and discussions that they wanted to engage with. Each post was within a category, and it was possible to filter the posts further using standard search tools, enabling users to easily (re)locate the posts that were of interest to them.

*Moderating NINEtoFIVE*: The research team (who were also union members) were responsible for moderating NINEtoFIVE throughout the deployment. A total of six members were involved in the moderation process, but a minimum of 3 members (and an average of 4 members) were present at any single moderation meeting. For the trial we decided to hold face-to-face moderation sessions for consequential analysis of the sessions. The face-to-face meetings took place twice per day: 8am for a 9am release and 4pm for a 5pm release. During the meetings, each message and comment was read aloud by a moderator, and to safeguard against the risk of inappropriate editing had to be approved by all moderators present at the meeting. Any 'challenge' to the post required an explicit case for either editing or removal. A post required a majority approval to be published on the system. The priorities for moderators were preserving the anonymity of both the author of a message (post or comment) and its subject, as well as preventing abusive messages, as well as mitigating the risk of 'jigsaw' identification through typos and idiosyncrasies that are characteristic of a particular author. The moderation was 'light touch' and did not correct every minor mistake, so many common typos remained. To enhance transparency, we displayed a "moderated" tag on a published message if it had been edited. This ensured the *Safety* and *Civility* of discussions. Moderators did not actively shape discussions, and controversial posts were allowed as long as they conformed to the aforementioned guidelines. Once all messages had been assessed, the moderators created a summary of the moderation session which included the three most popular posts and any comments discussed, as well as information regarding messages that were modified or rejected (e.g. number of messages and reasons). During the deployment moderators rejected 3 posts and 4 comments that were deemed to pose a risk for the author (i.e. that they might be identifiable). Where messages were rejected on the grounds of breaching anonymity, these were unanimous decisions by a team of moderators (at least 3) that were made in a small number of cases where the editing of messages (to anonymize them whilst still preserving the essential meaning) required contextual knowledge that the moderators did not have. While the moderators' rationale for message modification/rejection is recorded, the original messages themselves are permanently deleted. This information was then posted on the platform under a dedicated Moderators category with the aim of presenting transparent moderation practices, thus avoiding an atmosphere of conspiracy [73] and supporting the *Validity* of the system.

## FINDINGS FROM THE DEPLOYMENT

### Data Analysis

We conducted two types of analysis. First, we present descriptive statistics providing an account of system usage and behavior during the deployment period. Second, we followed this with a thematic analysis of the discussion threads [8], with the aim of understanding the extent to which our discourse principles had been supported by the system. For the thematic analysis all threads and messages were annotated to indicate the relationship between individual messages, as well as to indicate any connection between the design principles and goals, and the substantive content of what was said.

### Activity Patterns

The system released employee contributions at the beginning and end of each workday and as a result it would have been expected for the majority of the activity to take place around these times. Indeed, comments on existing posts and threads were mostly submitted after a new batch of posts were released (i.e. 9am or 5pm), whereas posts on new topics were generally spread throughout the day, suggesting that they were submitted as and when problems or ideas arose. Our data (Figure **2**) demonstrates that most activity was logged between 9am and 11am, after staff had arrived in the office and after the first message release of the day. **Figure 2** also shows the number of posts and comments that were submitted for moderation during the full two-week deployment. While we observe a natural drop-off in messages submitted over the course of the deployment (novelty effect), a reasonable amount of activity was sustained to the end of the deployment. Importantly, the number of views on the system during the second week are comparable to those during the first week (see **Figure 3**), showing that although posting activity decreased towards the end users were still checking the system right until the end of the deployment. The noticeable dip in activity in the centre of the graph can be attributed to both the weekend and a national holiday. The overall ratio of the number of employees to all users in the department is 144 to 604 (24%). In addition to attracting user contributions and views throughout the deployment period, the substantive content of posts revealed the two types of conversations on the platform: one-off posts (mostly complaints, see *'Taking Things Offline…'* below) or sustained conversations. Throughout this period, employees contributed 36 posts and 149 comments: 15 posts attracted comments over three or more days (40.5% of posts) and 22 posts attracted 3 or more comments overall (59.4% of posts: see **Figure 4**). Posts with 3 or more comments had 5.75 votes on average against 2.9 votes for posts with less than 3 comments. This is a good indication of discussion quality, given that any flaming comments would not have passed moderation. Previous work has found that discussions focusing on a very specific problem and discussed at a deep level are rated higher [63], supplying further support for the quality of discussions on NINEtoFIVE. Of course, there were other posts where few or no comments were made.

### Civility, Candour and Robustness

One important consideration was the quality of discussion: what behavior did our system encourage from our participants? The constructiveness and politeness of a conversation plays an important role in a group's ability to sustain a discussion and reach a point of agreement [53, 57]. In our analysis, we identified 9 threads (posts and/or comments) that included 17 total cases of impoliteness, including one thread with four cases. In addition to the 6 moderated entries which were altered for having inappropriate language, there were 23 potentially uncivil entries out of 185 published entries (12%). In addition, entries were modified due to typos and grammatical mistakes (n=9), formatting issues (n=6) and for clarification (n=6). There was only one 'clear' case of an uncivil entry, where the comment was impolite *and* did not contribute anything substantive to the discussion ('gender pay inequality'):

> SDP: *"Well this was clearly written by a man."*

All of the messages that were identified by moderators as uncivil (including the above) were due to vulgarity, rudeness or implied stereotypes, rather than aggressive or abusive content. In practice, the moderators adopted a liberal approach, one which was sometimes questioned. For example, during the discussion of noisy open plan offices, one of the user's suggestions included a suggestion to change jobs rather than to complain about their workplace. This was immediately picked up by other users and even served as a base for questioning the moderation process.

We identified two key discussion patterns: the divulging of personal experiences and confrontational points of view. The discussion of (anonymized) personal experiences or points of view was common in the majority of threads (29/36). For instance, the following account was provided by one poster:

> CCS: *"Yes I didn't used to think this is needed (for all sorts of reasons) but my recent observations on how male recruits at the University (even at junior levels) are allowed to negotiate starting salaries while female recruits are told "that's the level like it or lump it!" has changed my mind.'*

In general, the comments indicated that many users did not hesitate to express their agreement or disagreement with peers, although it sometimes led to uncivil responses – especially in threads that were judged as impolite. However, even in these cases the discussions returned to a constructive flow following one or two rounds of postings. *Another quality of these conversations* - especially in threads about 'Gender Pay and Justice' or 'Building Living' - was frank and open discussion with sometimes clear confronting points of view and willingness to engage into debates, for example:

> KUR: *'This is a tad vague. I'm curious as to what the OP thinks is a discriminatory practice, the fact that they might think one or two men have been treated better than them doesn't mean there is actual discrimination'*

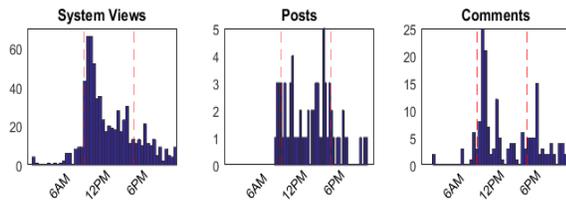

**Figure 2. Activity levels observed on the NINEtoFIVE system during the two–week deployment. The red lines indicate the times when posts and comments were released.**

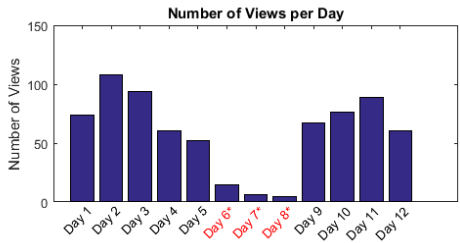

**Figure 3. Engagement levels (system views) observed on the system during the two–week deployment. Days 6–8 (in red text) were not working days.**

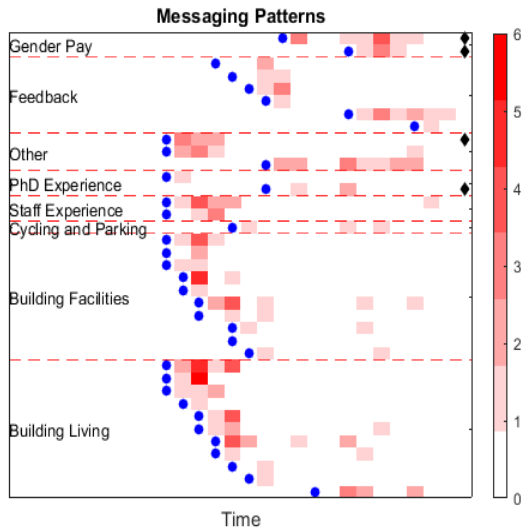

**Figure 4. Posting patterns on NINEtoFIVE, grouped by topic of discussion. Blue markers represent posts, whilst red colored squares represent the number of comments released in a moderation meeting.**

Overall, we did not find obvious disadvantages to discourse flow or continuity as a result of the *Anonymity* or *Slowness* of posting. Indeed, the participation and engagement was stable throughout the deployment and there was evidence of heated debates. However, the *Moderation* may well need to be adjusted, although it could be argued that robustness is an essential character of this type of discourse.

**Establishing Employee Voice**
*Workplace experience & discrimination*
One of the major discussion topics were posts related to organizational justice, and more specifically gender pay, equal opportunities and workplace discrimination. This topic amounted to 26 comments spread across four posts (making a total of 30 messages – 16.2% of all conversations) including some highly sensitive ones that highlighted the importance of the system's anonymity. By the end of the deployment, the primary discussion on gender pay was supported by additional threads concerned with gender discrimination and concerning allegations:

> UVW: 'my supervisor used to call for demonstrators, mail only to his male students and completely ignoring female students as if female students are not capable of demonstrating!'

In such message threads participants would raise issues supported by personal experiences, which were then later endorsed (or questioned) by comments and evidence from fellow colleagues.

> UPH: '@uvw that sounds like an obvious case of discrimination (which you might want to raise with the Union), ...'

In such cases we see tensions between ensuring the anonymity of the individuals while also allowing the collection of richer (other than text) forms evidence, for example through posting anonymized documents or recorded experiences. Also apparent was the lack of the system's affordances to put in motion a collaborative effort to address such issues, for example through generating alerts and reports for the management to consider such concerns particularly when workplace practices run the risk of contravening employment law. Where we did see more explicit calls for cooperative action was in response to a discussion about transparency in the workplace: '*The biggest failure is a lack of transparency – appointments and promotions seem based more on patronage, rather than merit.*' (ALS), whilst another explained that there should be more openness as to who is paid what. Perhaps unsurprisingly, pay disparity was considered a significant issue and provoked a significant number of posts. This was accompanied by a relatively lively discussion around work/life balance, with claims such as the following:

> USS: "I think the work/life balance is bad for so many phd students ... We need to stop boasting about how much time we spend on our work and discourage an environment where people work into the early hours of the morning to get things done it's not sustainable"

From the above, we can draw a number of conclusions. First, the process of this discussion was evolutionary: one claim of discrimination or unfairness led to other complaints, often on quite different topics. This resulted in not only the collection of a broad range of complaints but also their collation in ways that might not have been exposed by other means [74].

Second, the complaints themselves addressed sensitive issues that are seldomly voiced by employees in public fora. Even though this is the nature of discrimination complaints in general, NINEtoFIVE's *Anonymity* and *Moderation* allowed participants to openly air such complaints ensuring their *Safety* and *Egalitarianism* of voices. In a recent workplace survey, more than half of respondents had witnessed discrimination in the workplace and a less than a fifth reported the incident [32]. By discussing perceived discrimination on employee voice systems, employees are likely to improve the chances of these issues being reported to management, or at least raise awareness of such issues.

Finally, there were a number of people using the system as a form of social affirmation. The benefits of such social affirmation are well known in the health domain, where a range of eHealth platforms (such as [78]) exist to allow the sharing of sensitive health data. For the work context, the need for social affirmation might stem from a fear of reprisal on existing work communication platforms where the employees' identities are visible. We saw opportunities for such systems to accommodate such peer (and peer-to-peer) support mechanisms, while also underlining the importance of their sustainability and permanence.

*Taking things offline and responses to the system*
There were a set of posts that were explicitly about the move to the new building. Often these would involve a very specific complaint, either about the facilities themselves or people's use of them. These did not trigger much subsequent posting, possibly because they did not require further discussion, but were presumably meant to signal a problem to those who could potentially fix it, although for the purposes of this deployment, the complaints may have been unheard, as staff in the estates department were not active participants on this particular system deployment. This does raise an important issue: how can matters be taken matters *offline* in order to trigger action and adhere to *Efficacy*?

Some posts generated a 'call to action' that could be dealt with, by, and for, the staff themselves. For example, users organized physical wellbeing classes after extended online discussions on the system, followed by offline discussions facilitated by their voluntary de-anonymisation through a Google Form.

> VSC: 'yes yes yes! Let's do it! @weary tangerine armadillo if you can find out costs more of less then we could take it from there!'

In another example, users voluntarily contributed to a local food map that was then created and distributed (via leaflets and posters around the building) by the NINEtoFIVE team to support staff in their new location. Indeed, the wider consideration was described within discussion as being of fundamental importance, as expressed in the following representative example:

> SDR: "good work! Let's take this forward. Now, how do we get "out" of this platform and get something done. Clearly anonymity has its limitations. and it looks like we are far more interested to make stuff happen in here than just it being a talking shop."

We also found that not all staff within the building could access NINEtoFIVE online. A member of the research team encountered the cleaners having a chat over the posts on the system before work in the morning. We learned that they had been unable to access the system themselves due to their emails not initially being whitelisted, so they had asked someone with access to print the posts for their perusal. This suggests the implementation of *Control Access* principle could be refined. Another foible was concerning the legitimacy of the moderation process, with a proportion of the feedback received from users expressing concerns, including how the rotation of moderators should be publicized and be transparent:

> "The moderation is totally non-transparent. What posts or comments I am confident to submit depends on my knowledge on who is moderating it. How does one become a moderator anyway?..."

## DISCUSSION
### Balancing Anonymity, Slowness & Moderation

*Anonymity*: It is anonymity that allows employees to voice their concerns and address significant and sensitive issues openly, but anonymity can be a barrier to effective action. Even though in our deployment we did not observe any practical disadvantages for facilitating employee voice, it was clear that the anonymous nature of posts could lead to scepticism from management. Thus, the design of an effective employee voice system might require the provision of different levels of privacy, ranging from full anonymity, through partial anonymity and, in some cases, full identity disclosure. A possible mechanism for implementing this granular approach to anonymity is to design a support structure whereby participants could agree to identifying themselves in some kind of closed group, in order to engage in effective action – similar to that observed with the wellbeing classes subgroup. The transition from voice to action on our system was facilitated by the involvement of the trade union whose members would take up serious complaints. Ideally, if a claim (e.g. around discrimination) is raised on the platform, a third party (e.g. HR, or a trade union) should be able to take this forward, for example, by contacting the employee who raised the complaint. Based on this observations and findings we can suggest that *Anonymity* in itself does provoke more constructive and frank discussion without disadvantages on discourse flow or civility, however, it should be accompanied by the later step of actioning upon the discussion if necessary. It may be better to have a spectrum of possible anonymity configurations that improve the capacity to coordinate effective action.

*Slowness:* Most activity took place in the working week, and within the working day, although we note that the majority of activity was centered around two hours in the morning rather than impacting productivity throughout the day. Previous work has found that the more attentional switching that employees do during the day the less productive they feel at the end of the day [48]. Thus, the spike in activity during the morning, coupled with the fact that messages were only released twice per day (*Slowness* design response), meant attention switching was likely to be limited. While the evaluation of our deployment yielded some evidence that Slowness has contributed to the mitigation of flaming and other categories of aggressive behavior that is prevalent in anonymous systems, the real-world nature of our study and the absence of a control means we cannot make stronger claims to its impact. There was some decay in discourse continuation in some threads due to time (towards the end of deployment) or after resolutions. However, the smaller spike towards the end of the day might suggest that a 5pm release of posts is not necessarily appropriate for furthering discussion, thus suggesting the effect of the system can be improved by more carefully selecting the time for post-release. This might particularly be true in a less structured context (e.g. remote distributed teams), where a different view on slowness implementation might be required (e.g. delays).

*Moderation:* We have shown that the careful moderation of NINEtoFIVE can result in constructive discussions, unlike other high-profile anonymous platforms (e.g. [1, 34, 45, 68]). This is in line with previous observations where pre-moderated forums led to higher quality discussions [73] and less threatening posts [6]. Our analysis of activity patterns shows that the use of moderation did not negatively interrupt or influence discussions. However, the effort required to achieve this level of moderation (e.g. daily face-to-face meetings) is a concern for the scalability of the system. Much of the design space for efficient moderation remains to be explored. This includes: the use of remote moderation meetings; official moderator accounts that could be used to post when appropriate; and more detailed guidelines for both moderators (regarding the moderation process) and users (regarding civil discourse). Additionally, the moderation raised some questions from users. Thus, the transparency of the moderation process is an ongoing concern for both moderators and users: although, this must be seen in the context of online community governance where this continues to be a problem [54, 58]. This is in line with previous literature suggesting that the moderation process needs to be transparent in order to avoid an environment of conspiracy [73], yet measures were taken by the moderators to keep users informed of each session's outcomes via topic summaries and post edit/rejection stats (including the reason) [37]. Clearly this is not sufficient transparency to satisfy users, and future work will explore different methods for communicating this information to ensure continued user engagement. Moderation played a role in ensuring discussions remained civil, but we need to be careful that the moderation process does not interfere with the actual direction of conversations. More research is needed to understand whether moderating for *civility* can impact the likelihood of some conversations occurring or being avoided. Similarly, moderator team composition could in future have a wider constituency than union members and researchers.

**Facilitating Effective Employee Voice**
We have found that our system, developed based on identified design principles and goals, was able to effectively facilitate employee voice over the course of our trial period, generating a diverse range of constructive discussions in respect to issues that were often challenging, and unlikely to be fully discussed openly within the workplace. This success is evidenced by the volume and diversity of topics as well as evidence that controversial topics promoted a sustained discussion over the two-week period. In one sense, this demonstrates a significant degree of trust in the system, at least by those who chose to post: some of the comments could have had significant consequences for people if posted them under their own name (e.g. explicit claims of discrimination or overwork by their line managers or academic supervisors). The most notable point is that the system allowed the expression of concrete complaints of discrimination, which in turn could potentially be acted upon: a one-way feedback form would have been unlikely to produce these results. In relation to the issues of 'fear' and 'futility' raised earlier, we can evidence that some people overcame their fears in expressing genuine grievances, whilst recognising that the system does not guarantee that appropriate actions are taken in response to these concerns.

**Revisiting Employment and Rethinking Enrolment**
One unanticipated consequence of focussing our design activities on (and our partnership with) the trade union representing academic staff, research students and certain (higher pay grade) professional services staff, was that some other staff that were physically located in the department either did not have access to the system or faced significant challenges accessing it. One category of staff, external agency contractors, did not have university email addresses. Another category of staff, primarily central services staff (e.g. building managers) were not formally associated with the department or members of the academic trade union, and as such were not included in the list of users used to configure the system. In this respect, UK universities are not untypical of many large organizations for which the concept of employment (e.g. directly employed or subcontracted) and/or association with business units (e.g. member, business partner, service provider) is complex. While our multi-channel approach to engagement and promotion surfaced these issues, our failure to recognize this complexity from the outset, and to respond to it through the realisation of a truly inclusive enrolment process, highlights the fact that designing for *Validity* and *Egalitarianism* is in practice a greater challenge than we initially envisaged. A more comprehensive design process would have included engagement with a wider range of trade unions, both those recognized by the University and its subcontractors, as well as a more extensive consultation with staff 'on the ground' regarding the range of relevant stakeholders and how to access them.

**Acting on Issues Raised**
One aspect of NINEtoFIVE that requires significant further exploration is how it can inform and/or configure actionable outcomes [4, 13]. In our deployment, a number of specific actions resulted, including the initiation and scheduling of wellbeing classes, the creation of a map of local restaurants, and the resolution of some day-to-day equipment and infrastructure issues. In these cases, the uncontentious nature of the suggestions allowed participants to transfer the activities of planning and delivering change to traditional channels such a department mailing lists. However, self-organized collective action around more contentious issues is difficult to achieve, that is, how might occasional (anonymous) visitors to NINEtoFIVE become (anonymous) contributors and then (recognized) actors? Our participants made explicit requests that the 'evidence' from the system be considered by senior management to inform policy and practice. To this end we are currently considering how to best provide accountable data on system usage and content, recommendation for post-deployment activities (e.g. workshops based on such data), as well as more generalized guidelines and templates for an employee voice process (planning, deployment, analysis, action) that incorporates NINEtoFIVE.

**CONCLUSION**
In this paper we explored the concept of employee voice and identified the appropriate qualities for its facilitation: *Civility*, *Egalitarianism*, *Safety* and *Validity* of internal workplace conversations. We identified operationalized these principles as a set of design goals – *Assured Anonymity*, *Constructive Moderation*, *Adequate Slowness* and *Controlled Access* – for a digital system that can facilitate employee voice and provoke an establishment of horizontal peer-to-peer channels within an organization. During the subsequent deployment and evaluation of the developed anonymous system we showed that this design supports the aim of creating a trustful environment for constructive discussion and employee voice facilitation. Overall, our deployment of the system led to a diverse range of candid discussions around important workplace issues and produced some tangible change within the host organization.

**ACKNOWEDGEMENTS**
This work was partially supported by UKRI grants EP/M023001/1 and EP/L016176/1.